\documentclass[numreferences]{kluwer}

\usepackage{graphicx}

\newcommand{\bnvo}{$\beta-$Na$_{0.33}$V$_2$O$_5$}
\newcommand{\anvo}{$\alpha-$NaV$_2$O$_5$}
\newcommand{\cm}{cm$^{-1}$}
\newcommand{\figwidth}{0.95\textwidth}

\begin{document}
\begin{article}

\begin{opening}
\title{Charge and sodium ordering in \bnvo }
\runningtitle{Charge and sodium ordering in \bnvo} %
\subtitle{Raman and optical spectroscopy } \runningauthor{P.H.M. van
Loosdrecht {\it et al.}}
\author{P.~H.~M.~\surname{van~Loosdrecht}\footnote{\email{p.van.loosdrecht@phys.rug.nl}}}
\author{C.~N.~\surname{Presura}}
\author{M.~\surname{Popinciuc}}
\author{D.~\surname{van~der~Marel}}
\author{G.~\surname{Maris}}
\author{T.~T.~M. \surname{Palstra}} %
\institute{Material Science Center, University of Groningen,
             Nijenborgh 4, 9747 AG Groningen, The Netherlands}
\author{P.~ J.~ M.~\surname{van Bentum}}
\institute{University of Nijmegen, Toernooiveld, 6525 ED Nijmegen, The
Netherlands}
\author{H.~\surname{Yamada}}
\author{T.~\surname{Yamauchi}}
\author{Y.~\surname{Ueda}}
\institute{Institute for Solid State Physics, University of Tokyo, Japan}
\date{\today}

\begin{abstract}
Polarized Raman and optical spectra for the quasi-one dimensional metallic
vanadate \bnvo\ are reported for various temperatures. The spectra are
discussed in the light of the sodium and charge ordering transitions
occurring in this material, and demonstrate the presence of strong
electron-phonon coupling.
\end{abstract}
\end{opening}

\section{introduction}
The recent discovery \cite{yamada} of a clear metal-insulator transition
(MIT) in the vanadium bronze \bnvo\ has sparked a revival of interest in this
quasi one-dimensional metallic system. In addition to the MIT, \bnvo\
undergoes a structural sodium ordering transition at higher temperatures, a
magnetic transition at low temperatures, and a transition into a
superconducting state
under high pressure \cite{sctrans}. %
Several important aspects of this 1D material have remained unclear,
including the nature of the spin and charge excitations, and the relation
between the Na ordering and the MIT. This contribution presents a Raman and
optical study of \bnvo\ focusing on the various phase transitions and
electron phonon coupling in this compound.

\section{Structure and phase transitions}

\begin{figure}[p]
\begin{center}\includegraphics[width=\figwidth]{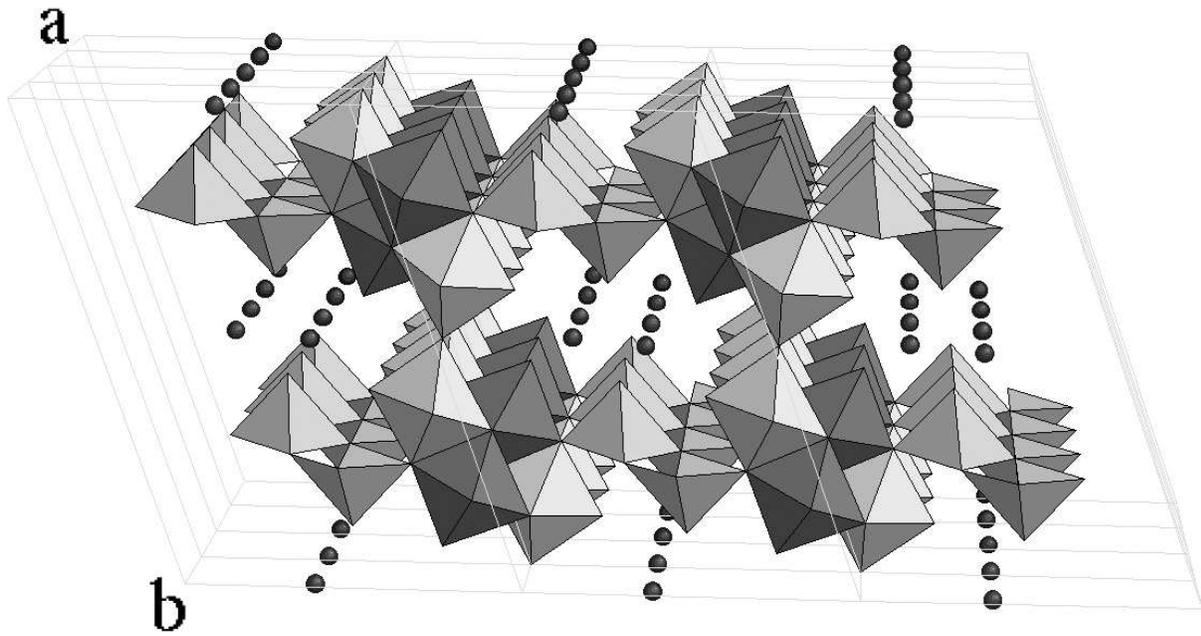} \end{center}
\caption{P.H.M. van Loosdrecht {\it et al.} \label{structure}}
\end{figure}

At room temperature \bnvo\ has a monoclinic structure (spacegroup C$^2_{3h}$,
a=10.08~\AA, b=3.61~\AA, c=15.44~\AA, $\beta$=109.6$^\circ$. )
\cite{wadsley,kobayashi}. The structure consists of zigzag double chains of
VO$_6$ octahedra, forming sheet by joining corners. These sheets are
separated by additional chains of double VO$_5$ trigonal bipiramids, giving
rise to unidirectional tunnels along {\bf b} (see Fig.\ref{structure}) in
which the Na ions are located. At a sodium stoichiometry of 0.33, there is
exactly one sodium atom per primitive cell. The sodium ion can occupy two
closely spaced positions, although simultaneous occupation of the two sites
is prohibited (the distance between the sites is 1.95~\AA). NMR experiments
\cite{maruyama} suggest that, at room temperature, the occupation of these
two sites is random giving rise to a disordered structure. At $T\simeq$240 K
a second order phase transition occurs in which the sodium atoms order in a
zigzag fashion along the unique axis accompanied by a doubling of the b-axis
\cite{yamada}. Two additional transitions occur at lower temperatures. At
$T\simeq$136 K a metal-insulator transition has been observed, in which the
unit cell undergoes and additional tripling along the b-axis \cite{tripling}.
It has been suggested that the charge ordering occurs due to a localization
of the charge on chains along the b-direction formed by the V$_1$ ions.
Finally, at $T\simeq$22 K there is a transition from the paramagnetic high
temperature state to a canted antiferromagnetic state \cite{vasiliev}

\section{Raman spectroscopy}

\begin{figure}[p]
\begin{center}\parbox[c]{0.5\textwidth}{\includegraphics[width=0.5\textwidth]{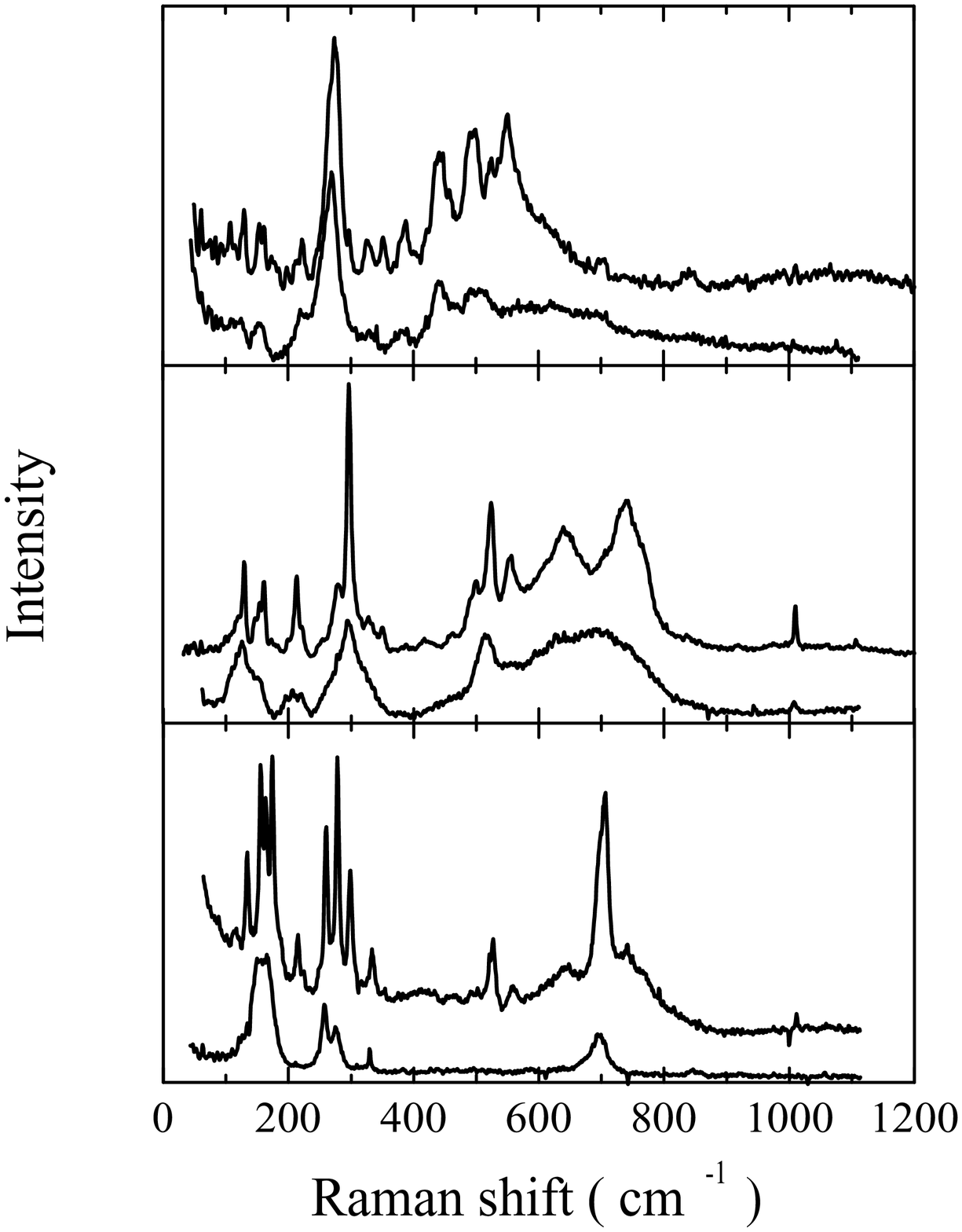}}
\parbox[c]{0.3\textwidth}{\includegraphics[width=0.3\textwidth]{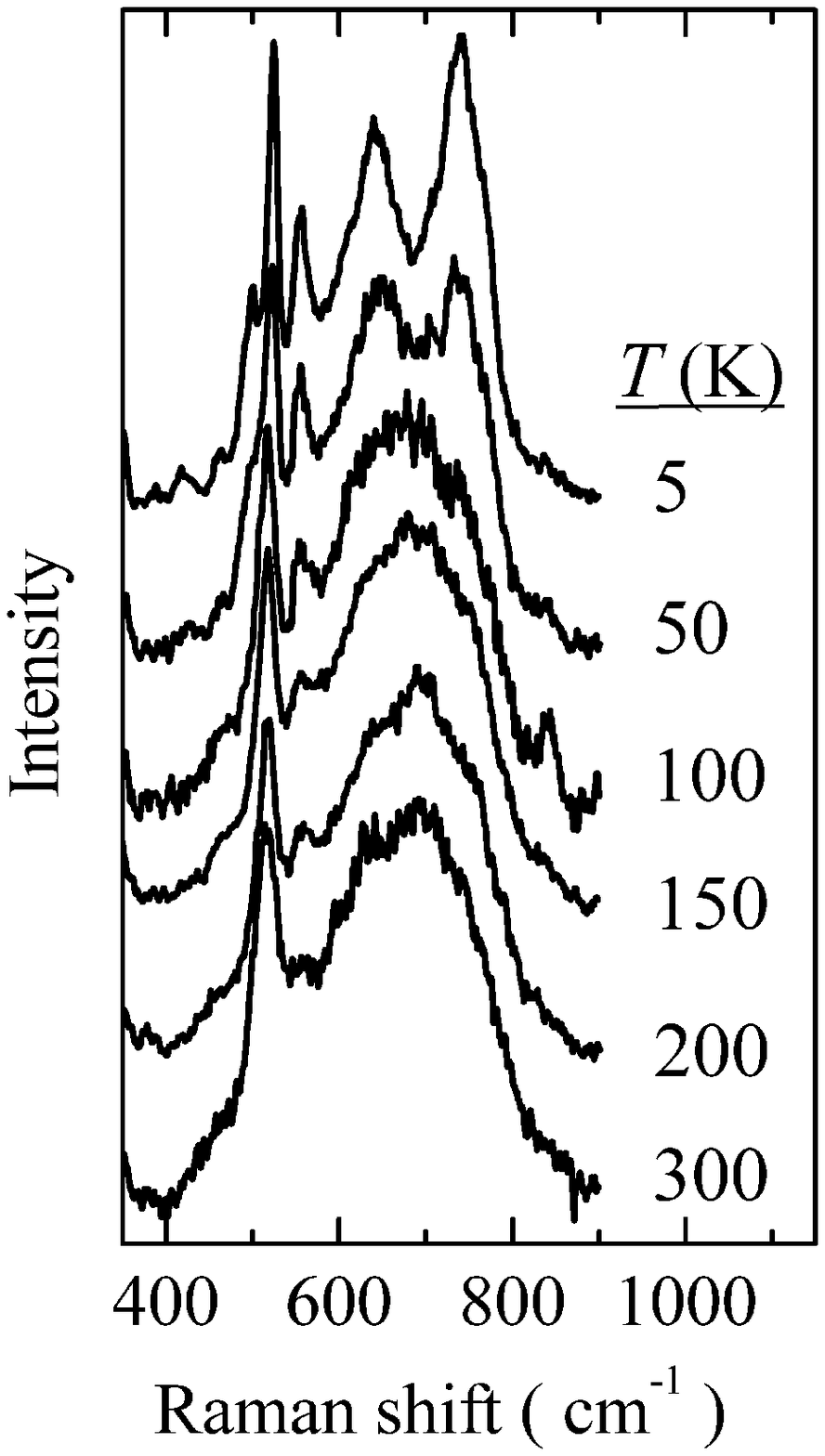}} \end{center}
\caption{P.H.M. van Loosdrecht {\it et al.} \label{raman}}
\end{figure}

The samples used in this study have been prepared as described in
 \cite{yamada}. Typical sizes are 5-6 mm along the b-direction and about
0.3-1~mm in the other two directions. For the Raman experiments, samples were
mounted in a flow cryostat (stabilization better than 1 K). Polarized spectra
have been recorded in a back reflection geometry using an Ar$^+$ ion laser
(514 nm) for excitation (power $<5$ mW, spot size 100 $\mu$m) and a state of
the art triple pass Raman spectrometer with diode array detection. Typical
spectra are shown in figure \ref{raman} (left part).

Due to the low symmetry and the large number of atoms in the unit cell, the
spectra show a large number of active phonon modes in all symmetries. What
immediately draws attention is the large width of most of the observed phonon
modes. In particular in the 400-800 \cm\ region, the line widths are 10-50
\cm. The active phonons here are expected to be vanadium-oxygen bending
modes. This strong broadening is believed to be due to a strong electron
phonon coupling, consistent with a polaronic picture of the electronic
properties of this material \cite{mott,presura}. The coupling is then a
result of the modulation of the hopping parameters for O(2p)-V(3d) hopping by
the phonons.

A group theoretical analysis show that the room temperature {\bf k}=0 optical
phonons can be classified as $\Gamma=11A_u+22B_u+20A_g+10B_g$, in which the
{\em gerade} modes are Raman active ($A_g$ in (aa),(bb), (cc), and (ac), and
$B_g$ in (ab) and (bc) polarization) and the {\em ungerade} IR active ($A_u$
for b polarization, $B_u$ for a and c polarizations). Consistent with the
crystal symmetry 16 modes are observed in $A_g$ symmetry, and 9 in $B_g$
symmetry. The missing modes might be to weak to be observed, but might also
escape detection due to near degeneracy.

Below the sodium ordering transition the crystal structure adheres to a
$C^5_{2h}$  \cite{tripling,gabi} symmetry. The symmetry analysis leads now to
a decomposition $\Gamma=65A_u+64B_u+66A_g+66B_g$\ for the optical modes.
Experimentally 55 modes are observed in $A_g$ symmetry, and 20 in $B_g$
symmetry. Below the charge ordering transition the unit cell triples, leading
to about 190 active phonons for each irreducible representation. Although a
few phonons seem to be activated below the metal insulator transition, the
spectra do not show this tripling. Again, this might be due to a lack of
scattering strength or near degeneracies.

The strongest changes in the spectra are observed below the metal insulator
transition at $T=136$~K. This is exemplified in figure \ref{raman} (right
panel). Apparently, the localization of the electrons leads to a decrease of
the coupling of the phonons to electronic excitations.

\section{optical spectroscopy}

\begin{figure}[p]
\begin{center}\includegraphics[width=\figwidth]{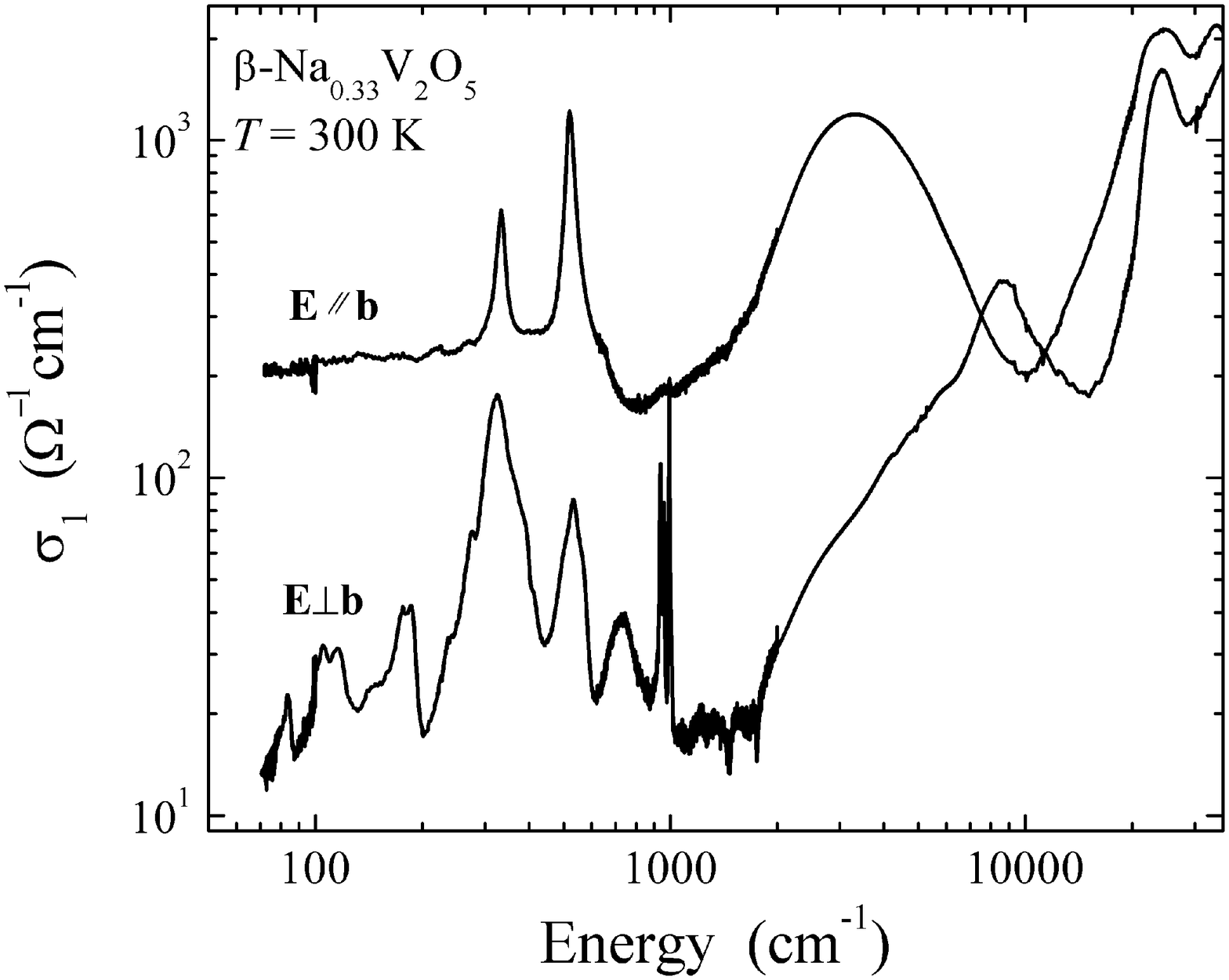} \end{center}
\caption{P.H.M. van Loosdrecht {\it et al.} \label{iroverview}}
\end{figure}

Temperature dependent optical constants of \bnvo\ in the NIR-UV range
(6000-35000 \cm) have been measured using ellipsometry on a surface
containing the b-axis. In addition the reflectivity spectra have been
measured as a function of temperature and polarization in the IR-MIR range
(20-6000 \cm). The optical conductivity of \bnvo\ has been determined by
combining the reflectivity data with the ellipsometric data and performing a
Kramers-Kronig analysis \cite{presura}. The room temperature results are
shown in figure \ref{iroverview}.

The strong absorption band observed above 10000 \cm\ in both spectra is due
to charge transfer excitations from the oxygen 2p to the vanadium 3d levels,
similar as has been observed in \anvo~ \cite{presura2}. The optical
conductivity $\perp$ b-axis shows an absorption band around 8000 \cm, again
similar to what is observed in \anvo. This band is probably due to a
bonding-antibonding transition on the V$-1$-O-V$_3$ bonds  \cite{presura}.

At lower energy the one dimensional metallic nature of \bnvo\ is clearly
demonstrated by the finite spectral weight for $E\rightarrow0$ for a
polarization along the b-axis, and the tendency to zero weight in the
perpendicular polarization. Indeed the data $\parallel$b may be fitted by a
Drude model, yielding an unscreened plasma frequency of $\simeq$3200~\cm, and
a scattering time $\tau\simeq$40~ps. This low energy response has been
attributed to mobile small polarons. The from the optical data estimated DC
conductivity is 200 $\Omega^{-1}$\cm, which compares well to the published DC
conductivity of 100 $\Omega^{-1}$\cm  \cite{yamada}. Another manifestation of
the polaronic nature of \bnvo\ is found in the characteristic \cite{polaron}
absorption peak around 3000~\cm.

\begin{figure}[p]
\begin{center}\includegraphics[width=0.85\textwidth]{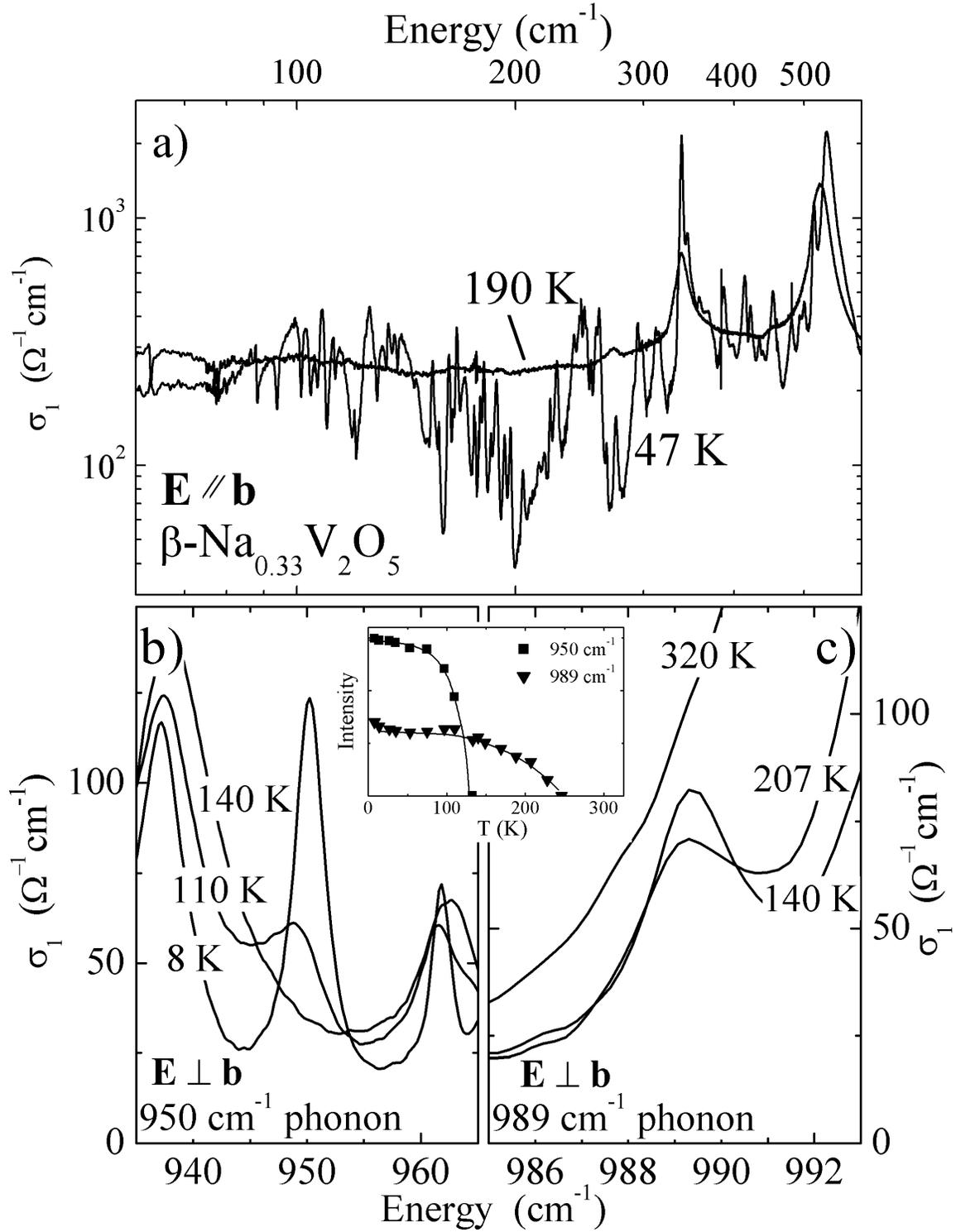} \end{center}
\caption{P.H.M. van Loosdrecht {\it et al.} \label{irdetail}}
\end{figure}

At room temperature one expects a total of 32 active phonon modes, 11 in
$A_u$ symmetry (polarization along b), and 22 in B$_u$ symmetry ($\perp$ b).
Along the insulating direction (figure \ref{iroverview}, lower curve) we
indeed observe 22 active phonon modes. In contrast, the metallic direction
(figure \ref{iroverview}, upper curve) exhibits only two strong phonon modes,
all other modes appear as strongly broadened weak features in the spectrum,
indicating again the importance of electron phonon coupling in this material.
Already below the sodium ordering transition one expects many new phonon
modes, though only few are actually observed (for an example see figure
\ref{irdetail} lower right panel). In contrast, the metal insulator
transition leads to a spectacular gradual appearance of a large number of
phonon modes (see figure \ref{irdetail}, upper and lower left panels). The
observed behavior is reminiscent of the observation of phase- or charged
phonons in several charge density wave materials, resulting from
electron-phonon interactions  \cite{rice}.

\section{conclusions}

The Raman and optical spectra presented in this contribution demonstrate in
many ways the importance of electron phonon coupling in \bnvo. The observed
phonon modes and symmetry are found to be in overall agreement with the phase
transitions in this material, although the appearance of the many modes in
the IR spectra along the metallic direction probably does not result from
symmetry breaking, but rather gain strength from electron phonon coupling in
the charge ordered phase. Finally, one interesting undiscussed aspect of the
optical data is the appearance of a continuum and Fano distortions of phonon
modes (see figure \ref{irdetail} upper panel). The origin of this low energy
continuum is presently unknown, but likely to be of electronic nature.

\acknowledgements This work was partially supported by the Dutch Foundation
for Fundamental Research on Matter (FOM) and by INTAS (99-155). %

\clearpage
\section*{Figure captions}
\begin{enumerate}
\item{Room temperature structure of \bnvo\ showing chains of V$_1$
(dark octahedra), V$_2$ (light octahedra), and V$_3$ ions (pyramids). }
\item{{\em left:} Polarized Raman spectra of \bnvo\ for $T=$15~K and
300~K. Upper panel A$_g$({\bf bb}), middle panel A$_g$({\bf
$\perp$b$\perp$b}), lower panel B$_g$({\bf b$\perp$b}) symmetry. %
{\em right:} Detailed temperature dependence of the 400-800 \cm\ region of
the A$_g$({\bf $\perp$b$\perp$b} Raman spectrum. In all panel the subsequent
curves have been given an offset for clarity. }
\item{Optical conductivity of
\bnvo\ at room temperature calculated from reflectivity spectra (FIR-MIR) and
ellipsometric data (MIR-UV). \cite{presura}. } %
\item{{\em upper panel: }
Optical conductivity of \bnvo\ along the metallic direction above and below
the charge ordering transition. {\em lower panels: } Phonon modes appearing
below the sodium (c) and charge ordering (b) transitions. The inset shows the
temperature dependence of the phonons shown in panels b) and c).}
\end{enumerate}

\clearpage

\clearpage

\clearpage

\clearpage

\begin{ao}\\
P.H.M. van Loosdrecht\\ University of Groningen \\Nijenborgh 4\\ 9747 AG
Groningen\\ The Netherlands
\end{ao}

\end{article}

\end{document}